\documentclass[sigconf,nonacm]{acmart}

\AtBeginDocument{%
  }

\usepackage{graphicx}
\usepackage{booktabs}
\usepackage{tabularx}
\usepackage{enumitem}
\usepackage{fontawesome5}
\usepackage{xcolor}

\begin{document}

\title{Paper Espresso: From Paper Overload to Research Insight}

\author{Mingzhe Du}
\email{mingzhe@nus.edu.sg}
\affiliation{%
  \institution{National University of Singapore}
  \city{Singapore}
  \country{Singapore}}

\author{Anh Tuan Luu}
\email{anhtuan.luu@ntu.edu.sg}
\affiliation{%
  \institution{Nanyang Technological University}
  \city{Singapore}
  \country{Singapore}}

\author{Dong Huang}
\email{dhuang@nus.edu.sg}
\affiliation{%
  \institution{National University of Singapore}
  \city{Singapore}
  \country{Singapore}}

\author{See-Kiong Ng}
\email{seekiong@nus.edu.sg}
\affiliation{%
  \institution{National University of Singapore}
  \city{Singapore}
  \country{Singapore}}

\renewcommand{\shortauthors}{Du et al.}

\begin{abstract}
The accelerating pace of scientific publishing makes it increasingly difficult for researchers to stay current.
We present \textbf{\textsc{Paper Espresso}}, an open-source platform that automatically discovers, summarizes, and analyzes trending arXiv papers.
The system uses large language models (LLMs) to generate structured summaries with topical labels and keywords, and provides multi-granularity trend analysis at daily, weekly, and monthly scales through LLM-driven topic consolidation.
Over 35~months of continuous deployment, \textsc{Paper Espresso} has processed over 13,300 papers and publicly released all structured metadata, revealing rich dynamics in the AI research landscape: a mid-2025 surge in reinforcement learning for LLM reasoning, non-saturating topic emergence (6,673 unique topics), and a positive correlation between topic novelty and community engagement ($2.0\times$ median upvotes for the most novel papers).
A live demo is available at \textcolor{blue}{\url{https://huggingface.co/spaces/Elfsong/Paper_Espresso}}.
\end{abstract}

\ccsdesc[500]{Information systems~Information extraction}
\ccsdesc[500]{Information systems~Summarization}
\ccsdesc[300]{Information systems~Web applications}

\keywords{paper summarization, trend analysis, knowledge discovery, large language models, research tools}

\maketitle

\section{Introduction}
\begin{figure}[ht]
  \centering
  \includegraphics[width=\linewidth]{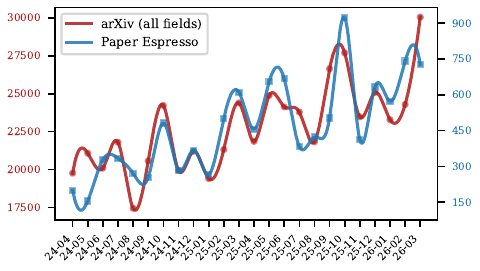}
  \caption{Monthly paper volume: arXiv total (red, left axis) vs.\ Paper Espresso (blue, right axis). Although Paper Espresso selects only community-trending papers ($\sim$2--3\% of arXiv), the two curves exhibit a consistent co-trend, confirming that the curated subset tracks the broader publishing rhythm.}
  \Description{A dual-axis line chart showing monthly submission counts from April 2024 to March 2026. The red line (left axis) represents total arXiv submissions ranging from 17,000 to 30,000. The blue line (right axis) represents HF Daily Papers curated by Paper Espresso ranging from 150 to 920. Both lines show an upward trend over time.}
  \label{fig:arxiv_vs_hf_monthly}
\end{figure}

The pace of scientific publishing now outstrips any individual researcher's capacity to stay informed. As shown in Figure~\ref{fig:arxiv_vs_hf_monthly}, arXiv alone receives nearly 30,000 submissions per month~\cite{arxiv_stats}, with no sign of deceleration.
This creates an acute \emph{information asymmetry}: the collective frontier advances rapidly, yet each researcher's awareness lags behind, filtered through keyword alerts and social media curation. The cost is not merely inconvenience but redundant efforts, missed cross-pollination, and delayed adoption of methodological advances.
Existing platforms such as Semantic Scholar~\cite{semantic2015}, Papers with Code~\cite{paperswithcode2019}, and ArXiv Sanity~\cite{karpathy2016arxivsanity}, along with LLM-powered tools like PaSa~\cite{pasa2025}, LitLLM~\cite{litllm2024}, and ScholarCopilot~\cite{scholarcopilot2025}, address fragments of this problem (indexing, retrieval, or writing assistance) but remain fundamentally \emph{reactive}: they require researchers to already know what to look for. None provides \emph{proactive, continuous monitoring} that combines structured paper comprehension with temporal trend analysis.

\begin{figure*}[ht]
  \centering
  \includegraphics[width=\textwidth]{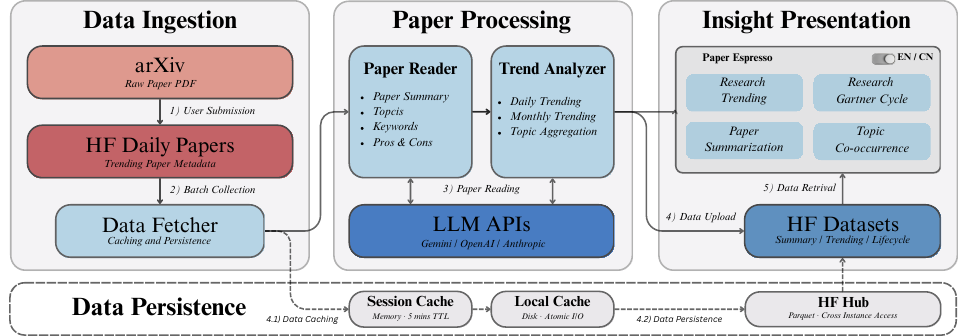}
  \caption{System architecture of \textsc{Paper Espresso}. The data ingestion layer fetches papers from the Hugging Face Daily Papers API and arXiv. The AI processing layer uses Google Gemini to generate structured summaries and trend analyses. The presentation layer provides an interactive Streamlit interface with multi-granularity browsing.}
  \Description{A system architecture diagram showing three layers: data ingestion from Hugging Face API and arXiv on the left, AI processing with Google Gemini in the middle, and the Streamlit web interface on the right. Arrows show data flow between components, with Hugging Face Hub datasets used for persistent storage.}
  \label{fig:architecture}
  \vspace{-0.5em}
\end{figure*}

We present \textbf{\textsc{Paper Espresso}}, an open-source system that continuously ingests community-validated trending papers, distills each into a structured summary, and proactively surfaces emerging research directions. Instead of indexing the full arXiv firehose, it targets the ${\sim}$2--3\% curated by the Hugging Face Daily Papers community and applies LLM-powered analysis to produce summaries, topical labels, keywords, and multi-scale trend reports. After 35~months of uninterrupted deployment, the system has grown into both a practical daily tool and a longitudinal observatory of the AI research landscape. It makes three contributions:
\begin{enumerate}[leftmargin=*, nosep]
\item \textbf{Open structured dataset.} We publicly release a structured dataset of LLM-generated paper summaries, topical labels, and keywords on Hugging Face~(13,388 papers, 6,673 topics, 51,036 authors), continuously updated via automated pipelines.
  \item \textbf{Multi-granularity trend analysis.} The system surfaces trending research directions at \textit{daily}, \textit{monthly}, and \textit{lifecycle} scales through LLM-driven topic consolidation, enabling researchers to track the evolving landscape without manual search.
  \item \textbf{Longitudinal empirical analysis.} Over 35~months of deployment, we reveal dynamics in the AI research landscape: a mid-2025 surge in \emph{reinforcement learning for LLM reasoning}, non-saturating topic emergence, a topic co-occurrence map exposing cross-cutting methodologies and emerging niches, and a divergence between topic frequency and engagement.
\end{enumerate}

\section{System Architecture}
\label{sec:architecture}
The system is organized as modular CLI-driven pipelines (daily, monthly, and lifecycle) backed by a Streamlit\footnote{\url{https://streamlit.io}} web frontend. All data is persisted to four public Hugging Face datasets in date-partitioned Parquet format, ensuring full reproducibility. As shown in Figure~\ref{fig:architecture}, the system comprises three layers: data ingestion, AI processing, and interactive presentation.

\subsection{Data Ingestion Layer}
Processing all ${\sim}$30,000 monthly arXiv submissions is neither feasible nor necessary; most researchers need only the high-impact subset. We therefore source papers from the Hugging Face Daily Papers API\footnote{\url{https://huggingface.co/papers}}, a community-curated feed where users upvote notable arXiv preprints. This yields a focused stream of ${\sim}$2--3\% of arXiv (Figure~\ref{fig:arxiv_vs_hf_monthly}), with upvote counts serving as a lightweight proxy for community attention. For each paper, the system captures the title, authors, abstract, arXiv identifiers, publication date, upvotes, and (when available) the full PDF for multimodal analysis.

\subsection{Paper Processing Layer}
The processing layer invokes LLMs via LiteLLM~\cite{litellm2025}, decoupling the data processing pipeline from any model provider.
A two-tier cache (local JSON checkpoints and remote Hub lookups) makes processing idempotent, so the pipeline skips already-summarized papers and resumes cleanly after any interruption.

\textbf{Paper Summarization.}
Each paper's title, abstract, and (when available) full PDF are sent as a single multimodal request. PDF grounding enables the model to capture methodological details beyond the abstract. The returned JSON contains: (1)~a concise summary (2--4 sentences), (2)~a detailed pros/cons analysis, (3)~open-vocabulary topic labels (2--3 free-form strings, not from a fixed taxonomy), and (4)~technical keywords (4--6 canonical terms, e.g., \textit{``LoRA,'' ``GRPO,'' ``DiT''}).

\textbf{Trend Analysis.}
\textit{Daily} reports distill the day's papers into dominant themes, a ranked topic list, and trending keywords. Open-vocabulary labeling naturally yields hundreds of fine-grained topics per month, far too many for direct browsing, so \textit{monthly} reports automatically consolidate them into ${\sim}$20 coherent clusters (e.g., \textit{``Multimodal LLMs'' and ``Vision-Language Models~(VLMs),'' $\to$ ``VLMs''}), with an explicit topic mapping back to the original per-paper labels. A bimonthly \textit{lifecycle} pipeline then classifies each topic into Gartner Hype Cycle~\cite{fenn2008mastering} phases using purely statistical indicators (Section~\ref{sec:analysis}), requiring no additional LLM calls.

\textbf{Bilingual Output.}
To serve both English-speaking and Chinese-speaking research communities, all LLM-generated fields are produced in both languages within a single call, eliminating a separate translation step. Chinese variants are stored alongside their English counterparts with a \textit{\_zh} suffix.

\subsection{Presentation Layer}
The web interface exposes three views. The \textbf{Daily} view lists papers sorted by upvotes, each rendered as a card with topic pills, the author list, and expandable TL;DR and pros/cons panels. The \textbf{Monthly} view deduplicates papers across the month and prepends an LLM-generated trend summary with ranked topics and keywords. The \textbf{Lifecycle} view presents a Gartner Hype Cycle chart alongside per-topic time-series of paper counts and proportions.

\section{Datasets}
\label{sec:dataset}

\textsc{Paper Espresso} publicly releases three complementary datasets on HF~Hub, continuously updated via the automated pipelines described in Section~\ref{sec:architecture}. All datasets are stored as date-partitioned Parquet files. Table~\ref{tab:stats} summarizes key statistics and Table~\ref{tab:schema} provides the complete field schema.

\paragraph{\textbf{Paper Summaries} (hf\_paper\_summary)}
Original paper metadata includes title, authors, abstract, publish date, upvotes, and full PDF.
LLM-generated fields include a summary (2--4 sentence TL;DR), a structured detailed analysis, open-vocabulary topics (2--3 labels), and keywords (4--6 terms). 

\paragraph{\textbf{Trending Reports} (hf\_paper\_daily/monthly\_trending)}
Each daily or monthly record contains a trending summary, ranked top topics, and trending keywords.
Monthly records additionally provide a topic mapping that traces each of the ${\sim}$20 consolidated clusters back to its constituent per-paper labels, enabling drill-down from coarse themes to individual papers.

\paragraph{\textbf{Lifecycle Snapshots} (hf\_paper\_lifecycle)}
Bimonthly snapshots store per-topic lifecycle classifications, monthly topic counts, and corpus-level statistics. These snapshots power the Hype Cycle visualization in the web interface and the lifecycle analysis in Section~\ref{sec:analysis}.

\begin{table}[t]
  \caption{Dataset statistics (May 2023 -- April 2026).}
  \label{tab:stats}
  \centering
  \small
  \begin{tabularx}{\linewidth}{Xlr}
    \toprule
    \textbf{Dataset} & \textbf{Records} & \textbf{Splits}         \\
    \midrule
    hf\_paper\_summary           & 13,388 & 733 days     \\
    hf\_paper\_daily\_trending   & 733    & 733 days     \\
    hf\_paper\_monthly\_trending & 34     & 34 months    \\
    hf\_paper\_lifecycle         & 18     & 18 bi-months \\
    \midrule
    \multicolumn{2}{l}{\textbf{Aggregate Statistics}} & \textbf{Count}  \\
    \addlinespace[2pt]
    \multicolumn{2}{l}{Unique papers}                         & 13,388  \\
    \multicolumn{2}{l}{Unique authors}                        & 51,036  \\
    \multicolumn{2}{l}{Unique Fine-grained topics}            & 40,565  \\
    \multicolumn{2}{l}{Unique Coarse-grained topics}          & 6,673   \\
    \multicolumn{2}{l}{Avg.\ fine-grained topics / paper}     & 3.03    \\
    \multicolumn{2}{l}{Avg.\ coarse-grained topics / month}   & 18.5    \\
    \multicolumn{2}{l}{Avg.\ upvotes}                         & 23.4    \\
    \bottomrule
  \end{tabularx}
\end{table}

\begin{table}[t]
  \caption{Field schema of the four released datasets.}
  \label{tab:schema}
  \centering
  \small
  \begin{tabularx}{\linewidth}{llX}
    \toprule
    \textbf{Field} & \textbf{Type} & \textbf{Description} \\
    \midrule
    \multicolumn{3}{l}{\textbf{Paper Summaries} (\textit{hf\_paper\_summary})} \\
    \addlinespace[2pt]
    paper\_id           & str  & arXiv identifier \\
    title               & str  & Paper title \\
    authors             & list & List of author names \\
    abstract            & str  & Original abstract \\
    upvotes             & int  & Community vote count \\
    published\_at       & date & Publication timestamp \\
    concise\_summary    & str  & TL;DR (avg.\ 551 chars) \\
    detailed\_analysis  & str  & Pros/cons analysis (avg.\ 1,827 chars) \\
    topics              & list & Fine-grained topic labels (avg.\ 3.03) \\
    keywords            & list & Extracted keywords \\
    \midrule
    \multicolumn{3}{l}{\textbf{Daily Trends} (\textit{hf\_paper\_daily\_trending})} \\
    \addlinespace[2pt]
    trending\_summary   & str  & Narrative overview of daily themes \\
    top\_topics         & list & Ranked dominant topics \\
    keywords            & list & Trending keywords of the day \\
    daily\_report       & str  & Human-readable daily report \\
    \midrule
    \multicolumn{3}{l}{\textbf{Monthly Trends} (\textit{hf\_paper\_monthly\_trending})} \\
    \addlinespace[2pt]
    trending\_summary   & str  & Monthly trend narrative \\
    top\_topics         & list & Consolidated topic clusters (15--20) \\
    topic\_mapping      & dict & Maps consolidated labels to originals \\
    monthly\_report     & str  & Detailed monthly analysis \\
    \midrule
    \multicolumn{3}{l}{\textbf{Lifecycle Snapshots} (\textit{hf\_paper\_lifecycle})} \\
    \addlinespace[2pt]
    lifecycle\_data      & dict & Per-topic phase, peak, slope, counts \\
    sorted\_months       & list & Ordered month labels in snapshot \\
    topics\_by\_month    & dict & Topic counts per month \\
    total\_by\_month     & dict & Total topic mentions per month \\
    n\_papers            & int  & Cumulative paper count at snapshot \\
    n\_months            & int  & Number of months in snapshot \\
    \bottomrule
  \end{tabularx}
\end{table}

\section{Empirical Analysis}
\label{sec:analysis}
Our analysis spans 35~months of deployment~(May 2023 to April 2026) and covers four dimensions: 
(1)~paper volume growth and community engagement patterns, 
(2)~topic distribution, temporal evolution, and co-occurrence structure dynamics.
(3)~topic lifecycle classification and velocity, and 
(4)~the relationship between paper novelty and community engagement.

\begin{figure*}[ht]
  \centering
  \includegraphics[width=\textwidth]{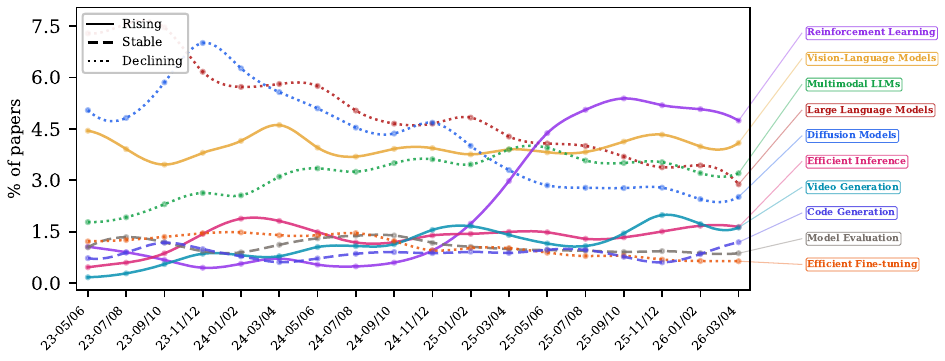}
  \caption{
    Bimonthly proportion (\%) of the top-10 research topics from May 2023 to March 2026, smoothed with a Gaussian kernel ($\sigma=0.8$) for visual clarity. 
    Trend arrows in the legend indicate each topic's recent trajectory.
  }
  \Description{A line chart showing bimonthly proportion trends for the top-10 topics. Each topic in the legend is annotated with a colored trend arrow: green for rising, gray for stable, red for declining.}
  \label{fig:topic_trends}
\end{figure*}

\subsection{Paper Volume and Community Engagement}
\begin{figure}[t]
  \centering
  \includegraphics[width=\columnwidth]{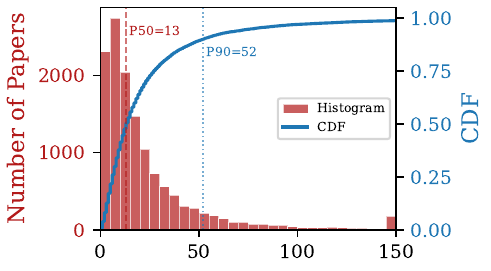}
  \caption{Community engagement distribution. The histogram~(red, left axis) shows a heavily right-skewed upvote distribution; the CDF~(blue, right axis) confirms that 50\% of papers receive $\le$13 upvotes and 90\% receive $\le$52.}
  \Description{A dual-axis chart with a red histogram of upvotes on the left axis and a blue CDF curve on the right axis. Vertical dashed lines mark P50=13 and P90=52.}
  \label{fig:engagement}
\end{figure}

Monthly intake grew from 259~papers in May~2023 to a peak of 923 in October~2025 (Figure~\ref{fig:arxiv_vs_hf_monthly}), averaging 18.8~papers on weekdays versus 3.3 on weekends, consistent with the academic publishing cycle. 
As shown in Figure~\ref{fig:engagement}, community upvotes are heavily right-skewed (skewness~$= 5.28$): the median paper receives 13~upvotes, yet the 90th~percentile reaches 52 and the maximum upvote is 664. This long tail means that upvotes carry genuine discriminative power: a uniformly distributed signal would make ranking meaningless, but the concentration of attention on the top~10\% of papers creates a clear separation between high-impact work and the majority, validating upvote-based ranking as a practical curation signal.

\subsection{Topic Landscape and Dynamics}

\paragraph{\textbf{Topic Distribution.}}
With an average of 3.03 topic labels per paper, the system produces 6,673 unique fine-grained topics across 13,388~papers (Table~\ref{tab:stats}).
Because labels are open-vocabulary (Section~\ref{sec:architecture}), lexically distinct but semantically equivalent labels (e.g., \textit{``VLMs''} vs.\ \textit{``Vision-Language Models''}) are counted separately; the monthly consolidation step merges such variants, reducing hundreds of labels to 15--20 coherent clusters (${\sim}$50:1 compression). Table~\ref{tab:top_topics} lists the five most frequent consolidated topics, which collectively cover over 56\% of all papers.

\begin{table}[ht]
  \caption{Top-5 consolidated research topics by paper count.}
  \label{tab:top_topics}
  \centering
  \small
  \begin{tabularx}{\linewidth}{Xrrr}
    \toprule
    \textbf{Topic} & \textbf{Count} & \textbf{\%} & \textbf{Cum.\%} \\
    \midrule
    Large Language Models           & 1,819 & 13.6 & 13.6 \\
    Vision-Language Models          & 1,598 & 11.9 & 25.5 \\
    Diffusion Models                & 1,514 & 11.3 & 36.8 \\
    Multimodal LLMs                 & 1,345 & 10.0 & 46.8 \\
    Reinforcement Learning          & 1,268 &  9.5 & 56.3 \\
    \bottomrule
  \end{tabularx}
\end{table}

\paragraph{\textbf{Topic Temporal Evolution.}}
Figure~\ref{fig:topic_trends} shows how topic dominance shifts over time. In early 2025, Large Language Models and Diffusion Models led the landscape. By mid-2025, Reinforcement Learning surged to the top, driven by rapid adoption of Group Relative Policy Optimization~(GRPO) and Reinforcement Learning with Verifiable Rewards~(RLVR) for LLM reasoning. VLMs remain consistently prominent, while Efficient Inference gains steady traction as deployment-oriented research matures.

\begin{figure}[t]
  \centering
  \includegraphics[width=\columnwidth]{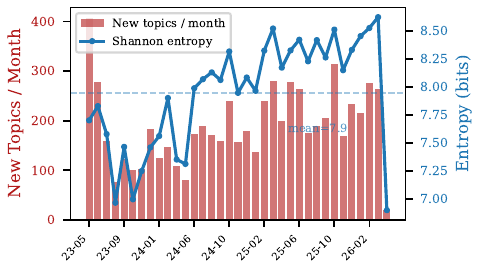}
  \caption{Topic emergence and diversity. Red bars show the number of new topics each month; the blue line tracks Shannon entropy of the monthly topic distribution, which remains flat around 7.9~bits, confirming sustained diversity.}
  \Description{A dual-axis chart with red bars showing 19--408 new topics per month on the left axis, and a blue line showing Shannon entropy fluctuating between 6.9 and 8.6 bits on the right axis with a dashed mean line at 7.9.}
  \label{fig:emergence}
\end{figure}

\paragraph{\textbf{Topic Emergence and Diversity.}}
As shown in Figure~\ref{fig:emergence}, new topics appear at a rate of 19--408 per month with no sign of saturation, while Shannon entropy $H = -\sum_i p_i \log_2 p_i$ over the monthly topic-frequency distribution remains stable around 7.9~bits (range 6.9--8.6). Together these indicate that the research frontier continues to diversify rather than collapsing toward a few dominant themes.

\begin{figure}[ht]
  \centering
  \includegraphics[width=\columnwidth]{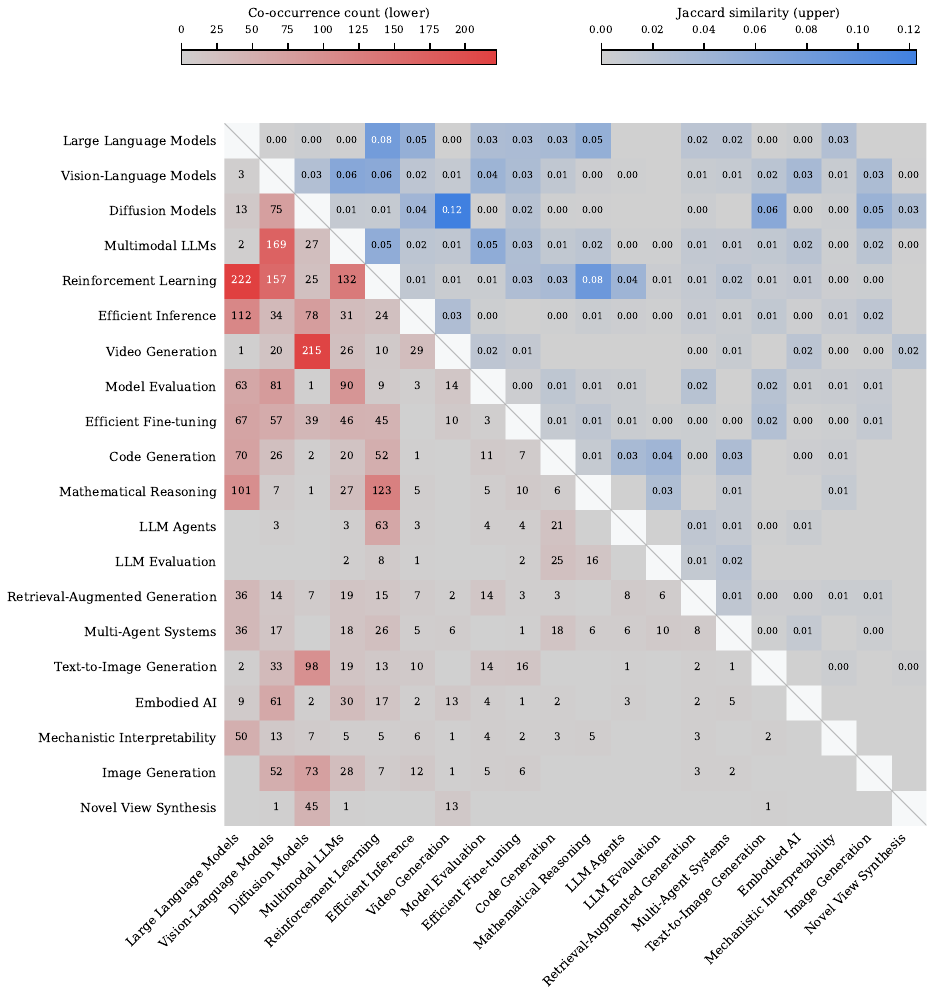}
  \caption{
    Co-occurrence heatmap for the top-20 topics. 
    The lower triangle shows raw co-occurrence counts (warm colors); 
    the upper triangle shows Jaccard similarity (cool colors), highlighting topic pairs that co-occur more than base rates.
  }
  \Description{A 20x20 split heatmap. The lower triangle uses a yellow-to-red colormap for co-occurrence counts (0--200). The upper triangle uses a yellow-to-blue colormap for Jaccard similarity (0--0.14). Two horizontal colorbars at the bottom label each scale.}
  \label{fig:cooccurrence}
\end{figure}

\paragraph{\textbf{Topic Co-occurrence.}}
Figure~\ref{fig:cooccurrence} shows raw co-occurrence counts (lower triangle) and Jaccard similarity $J = |A \cap B|/|A \cup B|$ (upper triangle) for the top-20 topics. 
Raw counts reflect absolute volume but are biased toward frequent topics; Jaccard normalizes by union size, revealing whether two topics co-occur more than their individual base rates would predict. 
Three patterns emerge:
(1)~\emph{RL as cross-cutting methodology}: Reinforcement Learning has the highest co-occurrence with LLMs~(215), VLMs~(152), Multimodal LLMs~(132), and Mathematical Reasoning~(123), permeating nearly every major direction.
(2)~\emph{Generative-vision cluster}: Diffusion Models pairs strongly with Video Generation~(197) and Text-to-Image~(71), with the Diffusion--Video pair also showing the second-highest Jaccard~(0.13), reflecting genuine technical coupling.
(3)~\emph{Frequency is not affinity}: the top-count pair (RL + LLMs, 215) has only moderate Jaccard~(0.09) because both topics are individually common, whereas Embodied AI and Vision-Language-Action Models share the highest Jaccard~(0.14) from just 50 papers, exposing a tightly coupled niche invisible to raw counts alone.

\begin{figure*}[ht]
  \centering
  \includegraphics[width=\textwidth]{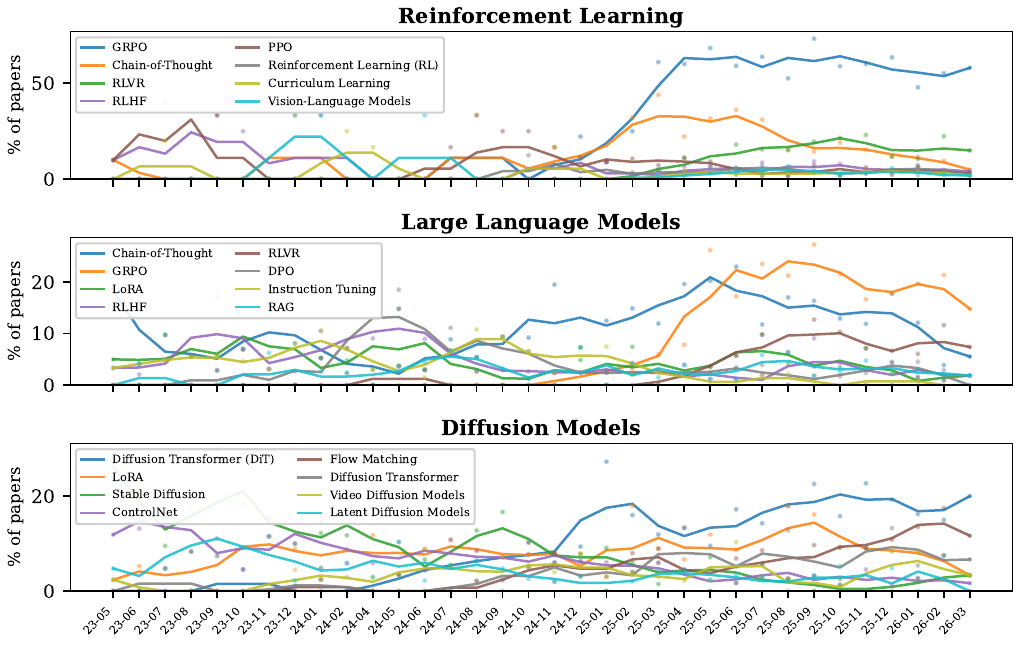}
  \caption{Keyword evolution within three major topics. Each line shows the percentage of papers (within that topic) mentioning a given keyword per month. Top: Reinforcement Learning shows a clear RLHF$\to$GRPO/RLVR transition. Middle: Large Language Models mirrors this shift. Bottom: Diffusion Models shows the UNet$\to$Transformer architectural migration.}
  \Description{Three vertically stacked line charts. Each panel tracks the top-8 keywords within one topic over time. The Reinforcement Learning panel shows GRPO surging to 65\% while RLHF declines. The LLM panel shows Chain-of-Thought and GRPO rising. The Diffusion Models panel shows DiT and Flow Matching replacing Stable Diffusion.}
  \label{fig:keyword_evolution}
  \vspace{-0.5em}
\end{figure*}

\paragraph{\textbf{Keyword Evolution.}}
Tracking keywords \emph{within} a topic reveals which specific methods drive its rise or fall. 
Figure~\ref{fig:keyword_evolution} traces the top-8 keywords for three major topics.
In \emph{Reinforcement Learning}, RLHF~\cite{ouyang2022rlhf} (${\sim}$25\% of RL papers in mid-2024) was rapidly displaced by GRPO~\cite{shao2024grpo} (${\sim}$65\% by early 2025) and RLVR~\cite{lambert2024rlvr}, marking a clear pivot from preference-based to verifiable-reward training.
\emph{Large Language Models} mirrors this shift: RLHF and DPO~\cite{rafailov2023dpo} declined while Chain-of-Thought~\cite{wei2022cot}, GRPO, and RLVR rose, signaling reasoning-oriented techniques as the new dominant paradigm.
In \emph{Diffusion Models}, the UNet-to-Transformer architectural migration is evident: Stable Diffusion~\cite{rombach2022stablediffusion} and ControlNet~\cite{zhang2023controlnet} faded while DiT~\cite{peebles2023dit} and Flow Matching~\cite{lipman2023flowmatching} gained steady traction.

\subsection{Topic Lifecycle}
We adapt the Gartner Hype Cycle~\cite{fenn2008mastering} to bibliometric data in order to characterize how research topics mature.
For every topic with at least 15~papers, we first compute its monthly proportion $p_t = c_t / N_t$, where $c_t$ is the number of papers assigned to the topic in month~$t$ and $N_t$ is the total number of topic assignments that month.
We then summarize each trajectory with five indicators: 
the \emph{peak proportion}~$p^{*}$ and the month at which it occurs; 
the \emph{current level}~$\bar{p}_{\text{cur}}$, averaged over the most recent 3 months; 
the \emph{decline ratio}~$\delta = \bar{p}_{\text{cur}} / p^{*}$, capturing how far the topic has fallen from its peak; 
the \emph{trend slope}~$\beta$, fit by Ordinary Least Squares~(OLS) over the last 6 months; 
and the \emph{recent fraction}~$\rho$, the share of a topic's papers published in the last 8 months.
Based on these indicators, each topic is assigned to one of five lifecycle phases:
\begin{enumerate}[leftmargin=*,nosep]
  \item \textbf{Innovation Trigger.} Newly emerging topics: active for $\le$8 months, or surging niches with $\rho > 0.60$ and $<$200~papers.
  \item \textbf{Peak of Inflated Expectations.} Topics near their all-time high ($\delta > 0.70$, peak within 6~months) or still rising strongly ($\beta > 0.001$, $\delta > 0.65$).
  \item \textbf{Trough of Disillusionment.} Topics well below peak with no sign of recovery ($\delta < 0.65$, $\beta \le 0.0003$), or actively declining ($\beta < -0.001$, $\delta < 0.75$).
  \item \textbf{Slope of Enlightenment.} Topics that have declined from peak but show renewed growth ($\delta < 0.65$, $\beta > 0.0003$).
  \item \textbf{Plateau of Productivity.} Mature, stable topics that match none of the above conditions.
\end{enumerate}

\begin{figure*}[ht]
  \centering
  \includegraphics[width=\textwidth]{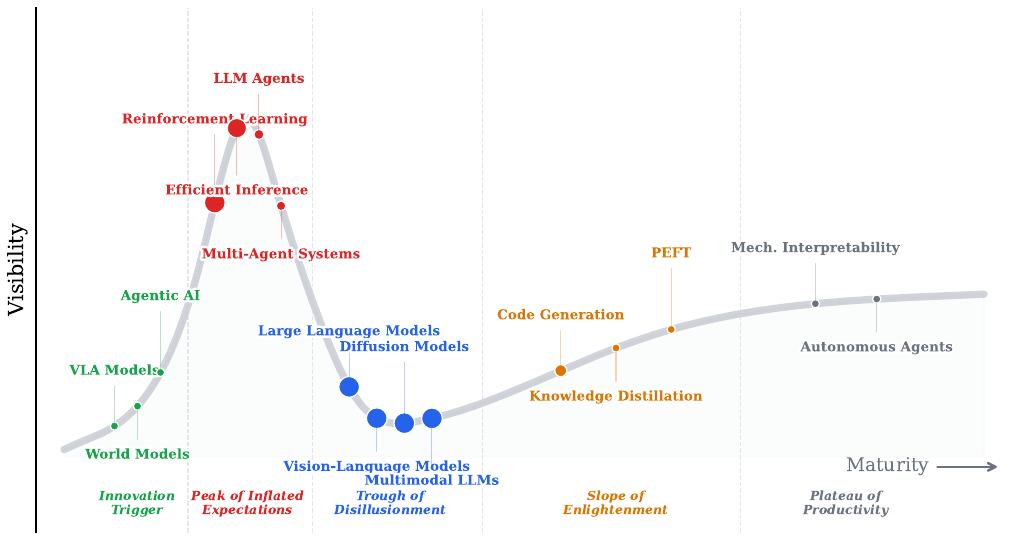}
  \caption{AI research hype cycle derived from 35~months of topic proportion time series. Topics are classified into five lifecycle phases based on peak timing, decline ratio, and recent trend slope. Dot size is proportional to total paper count.}
  \Description{A Gartner-style hype cycle curve with research topics positioned along it. Topics at the Peak include Reinforcement Learning and Efficient Inference; the Trough contains LLMs, VLMs, and Diffusion Models; the Slope of Enlightenment includes Knowledge Distillation and Prompt Engineering; the Plateau includes Mechanistic Interpretability.}
  \label{fig:hype_cycle}
\end{figure*}

Figure~\ref{fig:hype_cycle} maps notable topics to the lifecycle.
\emph{Reinforcement Learning}~\cite{sutton2018rl,du2025afterburner}, \emph{Efficient Inference}~\cite{zhou2024efficient,du2024mercury}, and \emph{LLM Agents}~\cite{yao2023react,huang2025nexus,ji2025verifiable} sit at the \textbf{Peak}, consistent with the mid-2025 surge in Figure~\ref{fig:topic_trends}.
\emph{LLMs}~\cite{zhao2023llmsurvey,ji2024cot}, \emph{VLMs}~\cite{zhang2024vlmsurvey}, and \emph{Diffusion Models}~\cite{yang2023diffusionsurvey} have entered the \textbf{Trough}, their proportional share declining even as absolute counts grow.
\emph{Knowledge Distillation}~\cite{hinton2015distilling} and \emph{Code Generation}~\cite{du2025codearena} occupy the \textbf{Slope of Enlightenment}, finding renewed applications after earlier decline, while \emph{Mechanistic Interpretability}~\cite{bereska2024mechinterp,wu2026deception} has reached a stable \textbf{Plateau}.
\emph{Vision-Language-Action Models}~\cite{kim2025openvla} and \emph{World Models}~\cite{ding2024worldmodels} appear at the \textbf{Innovation Trigger}, marking nascent research fronts.

\begin{figure}[t]
  \centering
  \includegraphics[width=\columnwidth]{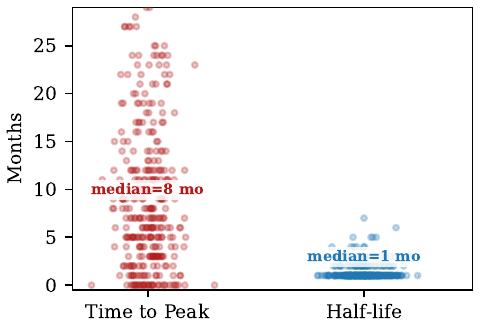}
  \caption{
    Topic velocity. 
    Topics take 8~months to peak (red) yet lose half their prominence within a single month (blue).
  }
  \Description{Side-by-side box-and-strip plots. Time to peak: median 8 months, wide spread. Half-life: median 1 month, tightly clustered.}
  \label{fig:topic_speed}
\end{figure}

\paragraph{\textbf{Topic Velocity.}}
For each topic with $\ge$15~papers and $\ge$4~active months, we measure \emph{time to peak} (months from first appearance to maximum proportion) and \emph{half-life} (months from peak to 50\% of peak). As shown in Figure~\ref{fig:topic_speed}, the contrast is stark: the median time to peak is 8~months, but the median half-life is just 1~month. AI research topics rise gradually yet decline abruptly, losing half their prominence within a single month of peaking. A few practically grounded topics resist this pattern, notably Instruction Tuning (7-month half-life), 3D Reconstruction~(6), and Efficient Inference~(4).

\subsection{Paper Novelty and Community Engagement}
We investigate whether papers with unusual topic combinations attract more community attention. 
For each paper with at least two topic labels, we define a novelty score as the negated mean Pointwise Mutual Information (PMI) across all co-assigned topic pairs: $\text{PMI}(t_i, t_j) = \log_2 \frac{P(t_i, t_j)}{P(t_i)\,P(t_j)}$, where co-occurrence probabilities are estimated from the full corpus with Laplace smoothing ($\alpha = 0.5$) for unseen pairs. Papers combining commonly co-occurring topics score low; those with unexpected pairings score high.

\begin{figure}[t]
  \centering
  \includegraphics[width=\columnwidth]{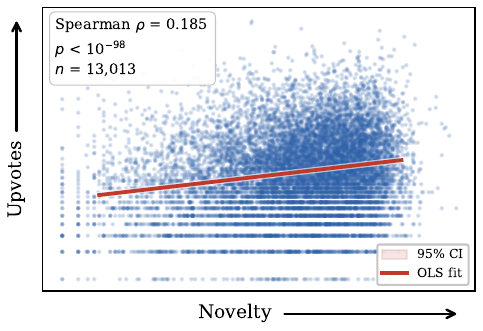}
  \caption{
    Novelty vs.\ engagement. 
    Papers with more novel topic combinations (higher scores) receive more upvotes.
  }
  \Description{Scatter plot of novelty score vs log-upvotes with a positive-slope OLS linear fit and 95\% confidence band.}
  \label{fig:novelty_engagement}
\end{figure}

As shown in Figure~\ref{fig:novelty_engagement}, novelty correlates positively with engagement. Frequency and engagement also diverge: Large Language Models is the most common topic, yet niche topics like Pre-training Strategies~(55), Computer Use Agents~(38), and Agentic Reasoning~(36) far exceed the global median of~14. Novelty and popularity thus carry complementary signals for paper recommendation.

\subsection{Takeaways}
\begin{enumerate}[leftmargin=*,nosep]
  \item \textbf{The AI research frontier is broadening, not converging.} New topics emerge at an undiminished rate (up to 408/month) while Shannon entropy remains stable (${\sim}$7.9~bits), indicating sustained diversification rather than consolidation around a few dominant themes. Researchers should actively monitor peripheral topics to avoid tunnel vision.

  \item \textbf{Topics peak slowly but fade fast.} The median topic takes 8~months to reach peak prominence yet loses half of it within a single month, making timely awareness critical. Systems that report trends only retrospectively (e.g., annual surveys) risk delivering insights after the window of opportunity has closed.

  \item \textbf{Novelty attracts attention.} Papers combining unexpected topic pairs receive $2.0\times$ the upvotes of those with conventional combinations. This suggests that the community rewards cross-pollination, and that recommendation systems should surface surprising intersections, not just popular categories.

  \item \textbf{Popularity and engagement are distinct signals.} The most \emph{frequent} topic~(LLMs, 13.6\% of papers) is far from the most \emph{engaging} per paper; niche topics such as Pre-training Strategies and GUI Agents draw $2$--$4\times$ higher median upvotes. Effective curation must weigh both volume and per-paper impact.
\end{enumerate}

\section{Related Work}
\label{sec:related}

\paragraph{\textbf{Academic Paper Discovery.}}
Semantic Scholar~\cite{semantic2015} offers large-scale indexing with AI-generated TLDRs~\cite{cachola2020tldr}, Papers with Code~\cite{paperswithcode2019} links papers to implementations, and ArXiv Sanity~\cite{karpathy2016arxivsanity} pioneered SVM-based personalized recommendation. LLM-era tools extend this landscape: PaSa~\cite{pasa2025} navigates citation graphs, LitLLM~\cite{litllm2024} applies RAG to literature reviews, and ScholarCopilot~\cite{scholarcopilot2025} fine-tunes a 7B model for citation-grounded writing. These systems are fundamentally \emph{reactive}, requiring users to know what to search for. \textsc{Paper Espresso} fills a different niche: \emph{proactive daily monitoring} that combines structured summarization with temporal trend analysis, so researchers discover what matters without issuing a query.

\paragraph{\textbf{Scientific Document Summarization.}}
Prior work ranges from discourse-aware attention models~\cite{cohan2018discourse} and extreme summarization~\cite{cachola2020tldr} to LLM-based scholarly review~\cite{liang2024llmreview}, with recent surveys charting this evolution~\cite{zhang2025textsumsurvey}. Unlike free-form summarizers, \textsc{Paper Espresso} produces \emph{structured} JSON output (summaries, pros/cons, topics), enabling programmatic filtering and aggregation.

\paragraph{\textbf{Research Trend Analysis.}}
Classical approaches include LDA~\cite{blei2003lda} for topic modeling, VOSviewer~\cite{vaneck2010vosviewer} for bibliometric mapping, and CiteSpace~\cite{chen2006citespace} for citation burst detection. Neural topic models such as BERTopic~\cite{grootendorst2022bertopic} and its temporal extension BERTrend~\cite{bertrend2024} offer embedding-based alternatives. Our system takes an orthogonal approach: instead of post-hoc analysis, it uses LLMs for \emph{real-time} topic labeling and consolidation as papers are published, producing human-readable trend reports within hours.

\section{Conclusion}
\textsc{Paper Espresso} is an open-source system that converts the daily stream of AI papers into structured summaries and multiple granularity trend reports. 
Analysis over 35~months reveals non-saturating topic emergence (6,673 unique labels), rapid topic decay (median half-life of one month), and a positive novelty-engagement effect ($2.0\times$ median upvotes for unconventional topic combinations). All code, data, and a live demo are publicly available.


\bibliographystyle{ACM-Reference-Format}
\bibliography{references}

\end{document}